\documentclass[aps,notitlepage,twocolumn,showpacs,preprintnumbers,floatfix,tightenlines, longbibliography, nofootinbib]{revtex4-1}
\usepackage{mathrsfs}
\usepackage{amssymb}
\usepackage{amsmath}
\usepackage{graphicx}
\usepackage{amsfonts,amsbsy}
\usepackage{nicefrac}
\usepackage{slashed}
\usepackage{xcolor}
\usepackage[breaklinks,colorlinks,urlcolor=blue,citecolor=citcolor,linkcolor=lcolor]{hyperref}
\usepackage{array}
\usepackage{multirow}
\usepackage{siunitx}
\usepackage{MnSymbol}
\usepackage[normalem]{ulem} 

\definecolor{lcolor}{rgb}{0.5,0,0}
\definecolor{citcolor}{rgb}{0,0.3,0.0}
\definecolor{ao(english)}{rgb}{0.0, 0.5, 0.0}
\definecolor{RoyalBlue}{HTML}{0071BC}

\newcommand{\mrm}{\mathrm}

\newcommand{\dA}{d_{\mrm{A}}}

\newcommand{\CA}{C_{\mrm{A}}}
\newcommand{\NC}{N_\mathrm{c}}

\newcommand{\CR}{C_{\mrm{R}}}
\newcommand{\Ceqappr}{C_{\mrm{eq}}^{\mrm{appr.}}}
\newcommand{\Cisoappr}{C_{\mrm{iso}}^{\mrm{appr.}}}

\newcommand{\fig}{Fig.~}

\newcommand{\vb}{\vec}
\renewcommand{\vec}[1]{\mathrm{\mathbf{#1}}}
\newcommand{\dd}[2][]{\mathrm d^{#1}{#2}\,}

\newcommand{\nmax}{n_{\mathrm{max}}}

\newcommand{\RE}{\mathrm{Re}}
\newcommand{\IM}{\mathrm{Im}}

\newcommand{\tform}{t^{\mathrm{form}}}

\newcommand{\bmax}{b_{\mathrm{max}}}
\newcommand{\bmin}{b_{\mathrm{min}}}
\DeclareSIUnit\c{c}

\begin{document}

\title{Gluon splitting rates in an anisotropic plasma in the AMY formalism}
\author{Florian Lindenbauer} 
\email{flindenb@mit.edu}
\affiliation{Institute for Theoretical Physics, TU Wien, Wiedner Hauptstraße 8-10, 1040 Vienna,
Austria}
\affiliation{MIT Center for Theoretical Physics -- a Leinweber Institute, Massachusetts Institute of Technology, Cambridge, MA 02139, USA}

\preprint{MIT-CTP/5920}

\begin{abstract}
We introduce a novel numerical method to obtain the gluon splitting rates in an anisotropic QCD plasma in the AMY formalism, suitable for an anisotropic collision kernel. The method extends previous works by decomposing the additional angular information into Fourier modes, resulting in a significantly larger system of differential equations to solve numerically. It is then tested by calculating the rates for a simple anisotropic model for the collision kernel, which is compared to a thermal system.
Remarkably, the obtained rates can be well-approximated by the rates calculated from an angular-averaged collision kernel, while still deviating significantly from equilibrium. An isotropic model for the collision kernel that is commonly used in QCD kinetic theory simulations, and which relies on the infrared temperature $T_*$ and an effective Debye mass, leads to rates that significantly deviate from the nonequilibrium rate, particularly at smaller parton energies.
\end{abstract}

\maketitle

\section{Introduction}
The quark-gluon plasma is a state of hot QCD matter characterized by deconfined quarks and gluons. While it may have existed in the earliest instances of our universe, it can nowadays be created experimentally in heavy-ion collisions, which are currently performed at the Relativistic Heavy Ion Collider (RHIC) and the Large Hadron Collider (LHC).
These experiments create an initially far-from-equilibrium plasma of deconfined quarks and gluons, which quickly thermalizes \cite{Busza:2018rrf}. 
Measuring hard probes such as jets, highly energetic particles created in the initial collision, provides a possible avenue to experimentally access properties of this deconfined medium \cite{Qin:2015srf, Apolinario:2022vzg}.

While traversing the medium, these energetic partons receive transverse momentum kicks, opening the phase-space for inelastic gluon emission processes, which dominate jet energy loss.
Significant effort has been devoted to developing the theory and framework to calculate such a splitting process \cite{Zakharov:1996fv, Zakharov:1997uu, Zakharov:1998sv, Baier:1996kr, Baier:1996sk, Baier:2000mf, Gyulassy:1999zd, Gyulassy:2000er, Gyulassy:2003mc, Wiedemann:2000za, Salgado:2003gb, Arnold:2002ja, Djordjevic:2008iz, Arnold:2008iy, Arnold:2008zu, Caron-Huot:2010qjx, Blaizot:2012fh, Mehtar-Tani:2019tvy, Mehtar-Tani:2019ygg, Barata:2021wuf}. While this process is theoretically often treated in the simplified case of a static or thermal medium, there has been recent progress to generalize the framework to include anisotropies or flowing media \cite{Romatschke:2004au, Romatschke:2006bb, Dumitru:2007rp, Hauksson:2021okc, Sadofyev:2021ohn, Andres:2022ndd, Barata:2022krd, 
Barata:2023qds, Kuzmin:2023hko, Barata:2024xwy}.

The fundamental quantity to describe this splitting process is the
\emph{dipole cross section} $C(\vb b)$,
which is often treated in a simple (harmonic) approximation, allowing for analytical results for the splitting rates and emitted energy spectra. The splitting rates beyond this approximation, but with the additional approximation of an infinite medium, have been obtained by Arnold, Moore, and Yaffe \cite{Arnold:2002ja}, and implemented in QCD kinetic theory simulations \cite{Arnold:2002zm, AbraaoYork:2014hbk, Kurkela:2014tea, Kurkela:2015qoa, Kurkela:2018oqw, Kurkela:2018xxd, Du:2020dvp, Du:2020zqg, kurkela_2023_10409474, Boguslavski:2024kbd, BarreraCabodevila:2025ogv}. However, in these QCD kinetic theory simulations, a simple isotropic approximation is used for the dipole cross section, or collision kernel $C(\vb q_\perp)$.

While the rate has recently been obtained for a specific model of an anisotropic collision kernel perturbatively for small anisotropies \cite{Hauksson:2023dwh}, this paper describes a general numerical method for obtaining the rate for a general anisotropic kernel and dipole cross section $C(\vb b)$. 
The numerical method is then tested using a simple model for this cross section, and the results are compared with a corresponding thermal model and with an isotropic approximation typically used in QCD kinetic theory simulations. Both the code and implementation used in this paper are publicly available \cite{lindenbauer_ratesscript}.

\section{AMY rates}
Let us start by discussing the formalism within which the gluon splitting rate is obtained.
For the gluon splitting process $g\to gg$, the rate
is given by \cite{Arnold:2002ja, Arnold:2002zm}
\begin{align}
    \gamma=\frac{p^4+p'^4+k'^4}{p^3p'^3k'^3}\frac{\dA\alpha_s}{2(2\pi)^3}\int\frac{\dd[2]{\vb h}}{(2\pi)^2}2\vb h\cdot \mathrm{Re} \vb F, \label{eq:gammarate}
\end{align}
where $\vb F$ is the solution to the integral equation
\begin{align}
    2\vb h &= i\delta E(\vb h)\vb F(\vb h)+\frac{1}{2}\int\frac{\dd[2]{\vb q_\perp}}{(2\pi)^2}C(\vb q_\perp)\label{eq:integralequation-amy}\\
    &\times \left[(3\vb F(\vb h)-\vb F(\vb h-p\vb q_\perp)-\vb F(\vb h-k\vb q_\perp)-\vb F(\vb h+p'\vb q_\perp)\right],\nonumber
\end{align}
with $\delta E(\vb h)=m_D^2/4\times (1/k+1/p-1/p')+h^2/(2pkp')$.
This expression for the rate is valid for all gluon energies, but was derived with the assumption of an infinite medium and that the collision kernel
\begin{align}
\begin{split}
    C(\vb q_\perp)&=g^2\CA\int\frac{\dd Q^0\dd Q^\parallel}{(2\pi)^2}2\pi\delta(v_{\hat n}\cdot Q)\\
    &\qquad\times v_{\hat n}^\mu {v_{\hat n}}^\nu\llangle A_\mu(Q)[A_\nu(Q)]^*\rrangle 
    \end{split}\label{eq:collisionkernel-formula-from-correlator}
\end{align}
does not change significantly (i.e., is constant) during the formation time $\tform\sim\sqrt{\omega/\hat q}$ of a splitting process. Other processes such as gluon radiation off a quark, $q\to qg$, or pair creation $g\to q\bar q$ can be obtained similarly with different color factors \cite{Arnold:2002ja, Arnold:2002zm}. The method introduced in this paper can thus be applied to these processes as well.
The double brackets $\llangle A_\mu(Q)[A_\nu(Q)]^*\rrangle$ denote the mean square fluctuations of the background gauge fields and is given by the (Fourier transformed) Wightman correlator.
Despite these approximations, these rates constitute an important ingredient to QCD kinetic theory, where they enter in the inelastic collision term \cite{Arnold:2002zm}.

Previously in the literature, numerical evaluations of this rate only considered isotropic collision kernels (or, equivalently, isotropic dipole cross sections) \cite{Andres:2020vxs, Andres:2023jao, Caron-Huot:2010qjx, Moore:2021jwe, Schlichting:2021idr, Yazdi:2022bru, Modarresi-Yazdi:2024vfh}, or solved the rate equation \eqref{eq:gammarate} perturbatively around isotropy \cite{Hauksson:2023dwh}. The method described here can be used to obtain the rate \eqref{eq:gammarate} for a general anisotropic collision kernel $C(\vb q_\perp)$.

For a numerical solution, it is often convenient to solve the integral equation \eqref{eq:integralequation-amy} in impact parameter space (see, e.g., \cite{Aurenche:2002wq, Moore:2021jwe, Hauksson:2023dwh}), where the relevant quantity is the dipole cross section
\begin{align}
    C(\vb b)=\int\frac{\dd[2]{\vb q_\perp}}{(2\pi)^2}(1-e^{i\vb b \cdot \vb q_\perp})C(\vb q_\perp), \label{eq:fouriertrafo}
\end{align}
which will also be the relevant quantity for the method described in Section \ref{sec:numerical-method}.

\section{Collision kernel in equilibrium and anisotropic models\label{sec:collkern-equ-anisotropic}}
Before the method is described in detail, let us discuss a simple model for an anisotropic collision kernel.

In thermal equilibrium, the collision kernel can be obtained as an infinite sum of modified Bessel functions \cite{Arnold:2008vd}.
At small $q_\perp$, it has the compact analytic limit
\begin{align}
    C(q_\perp)=\frac{\CR g^2m_D^2 T}{q_\perp^2(q_\perp^2+m_D^2)}.\label{eq:small-qperp-form-collkern-dipolecrosssectionsection}
\end{align}
Although this form is only valid for small $q_\perp$, it will be used here as a model for the collision kernel for general $q_\perp$, in particular entering the transformation to impact parameter space \eqref{eq:fouriertrafo}.
The Fourier transform \eqref{eq:fouriertrafo} can then be performed analytically,
\begin{align}
    \Ceqappr(b)&=\frac{C_Rg_s^2T}{2\pi}\left(\gamma_E+K_0(b m_D)+\log\frac{bm_D}{2}\right),\label{eq:Cb_eq_appr}
\end{align}
where $K_0$ is the modified Bessel function of the second kind with $\alpha=0$. Note that this is only an approximate form of the thermal dipole cross section, since it is obtained from taking the small-$q_\perp$ limit of the collision kernel \eqref{eq:small-qperp-form-collkern-dipolecrosssectionsection}, while the integral \eqref{eq:fouriertrafo} is over arbitrary values of $q_\perp$. Nevertheless, it is a convenient compact analytic expression, which will be used in the following.

Ref.~\cite{Hauksson:2023dwh} introduces a simple anisotropic model for the collision kernel, where the Debye mass $m_D$ in  Eq.~\eqref{eq:small-qperp-form-collkern-dipolecrosssectionsection} is endowed with an angular dependence,
\begin{align}
    m_D^2(\phi)=\left(1-\frac{2\xi}{3}+\xi\cos^2\phi\right)\bar m_D^2.\label{eq:debyemass-anisotropic-model}
\end{align}
This is motivated by the behavior of the screening mass of a squeezed thermal distribution function \cite{Romatschke:2003ms},
\begin{align}
    f(\vb p;\xi)=A(\xi)\,n_B(p_x^2+p_y^2+p_z^2(1+\xi)).
\end{align}
The distribution is characterized by an anisotropy parameter $\xi\in (-1,\infty)$ and a suitable normalization factor $A(\xi)$. In this paper, this factor is chosen such that the energy density
\begin{align}
    \varepsilon=\int\frac{\dd[3]{\vb p}}{(2\pi)^3}|\vb p|f(\vb p)
\end{align}
remains constant when varying the anisotropy parameter $\xi$, which leads to 
\begin{align}
    A(\xi)=\frac{2\sqrt{1+\xi}}{\int_{-1}^1 \dd{x}\sqrt{1-x^2+\frac{x^2}{1+\xi}}}.
\end{align}

The effective Debye mass $\bar m$ is obtained from the isotropic distribution $f_{\mathrm{iso}}(p)=A(\xi)n_B(p)$, 
such that $\bar m_D=\sqrt{A(\xi)}m_D.$

Plugging the anisotropic screening mass \eqref{eq:debyemass-anisotropic-model} into \eqref{eq:small-qperp-form-collkern-dipolecrosssectionsection} and expanding for small $\xi$, we obtain
\begin{align}
    C(\vb q_\perp)=\frac{\CR g^2\bar m_D^2T}{q_\perp^2(q_\perp^2+\bar m_D^2)}+\xi \frac{\CR g^2 \bar m_D^2T(3\cos^2\phi-2)}{3(\bar m_D^2+q_\perp^2)^2}+\mathcal O(\xi^2),
\end{align}
which can be analytically Fourier transformed,
\begin{align}
\begin{split}
    C(\vb b)&=\Ceqappr(b)\\
    &+\frac{\xi}{6(2\pi)^2}\CR g^2\pi T \Bigg[-1+b \bar m_D K_1(b \bar m_D)(1-3\cos(2\phi_b))\\
    &\qquad +6\left(\frac{2}{b^2 \bar m_D^2}-K_2(b \bar m_D)\right)\cos2\phi_b\Bigg]+\mathcal O(\xi^2).
\end{split}
\end{align}

It will be useful to also calculate the Debye mass $m_D$, which reads
\begin{align}
    m_D^2=4\lambda \int\frac{\dd[3]{\vb p}}{(2\pi)^3 |\vb p|}f(\vb p)=\frac{(m_D^0)^2A(\xi)\int_{-1}^1\frac{\dd{x}}{\sqrt{1-x^2+\frac{x^2}{1+\xi}}}}{2\sqrt{1+\xi}}. \label{eq:debyemass-calculated}
\end{align}
Similarly, the infrared temperature $T_\ast$ is given by
\begin{align}
    T_\ast&=\frac{2\lambda}{m_D^2}\int\frac{\dd[3]{\vb p}}{(2\pi)^3}f(\vb p)(1+f(\vb p))\\
    &=T\frac{4\NC\pi}{\int_{-1}^1\frac{\dd{x}}{\sqrt{1-x^2+\frac{x^2}{1+\xi}}}}\frac{8\zeta(3)+\frac{4}{3}(\pi^2-6\zeta(3))A(\xi)}{(2\pi)^3}.
    \label{eq:tstar-calculated}
\end{align}
Enforcing the energy density to remain constant implies that the parameter $T$ can be identified as the temperature of the thermal system with the same energy density as the nonequilibrium system (Landau matching).

The nonequilibrium model will also be compared with an isotropic model for the collision kernel, 
\begin{align}
	\Cisoappr(b)&=\frac{C_Rg_s^2T_\ast}{2\pi}\left(\gamma_E+K_0(b m_D)+\log\frac{bm_D}{2}\right),\label{eq:Cb_iso_appr}
\end{align}
with the Debye mass $m_D$ as defined in \eqref{eq:debyemass-calculated} and $T_\ast$ given by \eqref{eq:tstar-calculated}.

It should be emphasized that this simple anisotropic model is intended to illustrate the numerical method described in the following section. The method is more general and can be applied to any form of the collision kernel, as is done in the companion paper \cite{Altenburger:2025iqa}, where the nonequilibrium gluon splitting rates are obtained for a collision kernel obtained from a QCD kinetic theory simulation of the initial stages in heavy-ion collisions.

\section{Numerical method\label{sec:numerical-method}}
In this section, the numerical method employed in this paper is described in detail. Its implementation is publicly available \cite{lindenbauer_ratesscript}.

\subsection{Differential equation in impact parameter space}
The method described in this paper follows and generalizes the method outlined in Ref.~\cite{Aurenche:2002wq}.\footnote{More recently, a similar but slightly different method was described in Ref.~\cite{Moore:2021jwe} for a thermal system. This method uses the known analytic solution for the vacuum case, i.e., a vanishing collision kernel, to isolate its diverging contribution. The full thermal rate is obtained by solving a modified differential equation with the collision kernel entering as an inhomogeneity. However, for obtaining the inhomogeneous solution, the collision kernel, particularly its form and angular dependence, is relevant also at small $b$, which complicates setting the boundary conditions. In the method described in this paper, the boundary conditions can be imposed at small $b$ where the exact form of the collision kernel does not contribute.
} The integral equation~\eqref{eq:integralequation-amy} is solved in impact parameter space using the Fourier transformed quantities
\begin{subequations}\label{eq:app-fouriertransform-2d}
\begin{align}
    \vb F(\vb b)&=\int\frac{\dd[2]{\vb h}}{(2\pi)^2}e^{i\vb b\cdot\vb h}\vb F(\vb h),\\
    \vb F(\vb h)&=\int\dd[2]{\vb b}e^{-i\vb b\cdot\vb h}\vb F(\vb b), \\
    \tilde C(\vb b)&=\int\frac{\dd[2]{\vb q_\perp}}{(2\pi)^2}e^{i\vb b\cdot\vb q_\perp}C(\vb q_\perp).
\end{align}
\end{subequations}
In Eq.~\eqref{eq:integralequation-amy}, $\tilde C(\vb b)$ always appears as the difference
\begin{align}
    C(\vb b)=\tilde C(0)-\tilde C(\vb b),
\end{align}
which is introduced above as the \emph{dipole cross section}.
With the abbreviations
\begin{align}
    A&=\frac{im_D^2}{4p}\left(\frac{1}{z}+\frac{1}{1-z}-1\right),\\
    B&=\frac{i}{2pz(1-z)},
\end{align}
the integral equation \eqref{eq:integralequation-amy} can be written as
\begin{align}
    \begin{split}
    2\vb h&=\vb F(\vb h)(A+B\vb h^2)\\
    &+\frac{1}{2}\int\frac{\dd[2]{\vb q}}{(2\pi)^2}C(\vb q)
    \Big\{3\vb F(\vb h)-\vb F(\vb h-\vb q)\\
    &\qquad -\vb F(\vb h-z\vb q)-\vb F(\vb h-(1-z)\vb q)\Big\}.
    \end{split}
\end{align}
Note that $A, B \in\mathbb C$ are complex but purely imaginary such that $iA \in \mathbb R$, $A/B\in\mathbb R$.

Inserting now the Fourier transforms \eqref{eq:app-fouriertransform-2d}, we obtain in impact parameter space
\begin{align}
    (A-D(z,\vb b)-B\nabla^2)\vb F(\vb b)=-2i\nabla \delta^{(2)}(\vb b),\label{eq:app-impactparameterspaceequation2}
\end{align}
where the function $D(z,\vb b)$ is given by
\begin{align}
    D(z,\vb b)=-\frac{1}{2}\left(C(\vb b)+C((1-z)\vb b)+C((1-z)\vb b)\right).
\end{align}
Methods to solve this equation for isotropic $D(z, b)$ have been described in Refs.~\cite{Aurenche:2002wq, Moore:2021jwe}. 
The method described here generalizes them. In Section \ref{sec:special-case-isotropic}, the specialization to isotropic systems is revisited.

First, note that the delta function can be seen as imposing a boundary condition on $\vb F$. This can be seen by considering the differential equation obtained by including the most singular terms
\begin{align}
    -2i\nabla\delta(\vb b)=-B\nabla^2 F(\vb b).\label{eq:differentialequation-delta}
\end{align}
This is solved by
\begin{align}
	\lim_{b\to 0}\vb F(\vb b)=\frac{i}{B\pi}\frac{\vb b}{b^2},\label{eq:boundarycondition}
\end{align}
which is demonstrated in App.~\ref{app:delta}.
This boundary condition is used to solve Eq.~\eqref{eq:app-impactparameterspaceequation2} in the region of positive nonzero $b>0$.

Next, we redefine $\vb F=b {\vb g }=b(g_x, g_y)$, for which the differential equation becomes
\begin{align}
    \left[\frac{A-D(b,\phi_b)}{B}-\partial_{b}^2-\frac{3}{b}\partial_{b}-\frac{1}{b^2}-\frac{1}{b^2}\partial_{\phi_b}^2\right]\vb g(b,\phi_b)=0.
\end{align}
This redefinition has the advantage that the real part of the integral appearing in \eqref{eq:gammarate} can be obtained by taking the imaginary part of the function $\vb g$ evaluated at $b\to 0$.
\begin{align}
    \int\frac{\dd[2]{\vb h}}{(2\pi)^2}\RE\,\left(\vb h\cdot \vb F(\vb h)\right)&=
    \RE\,\left(-i\nabla\cdot\vb F(\vb b)\right)_{\vb b=0}\\
    &=\IM\,\left[\frac{g_x(0)}{\cos\phi_b}+\frac{g_y(0)}{\sin\phi_b}\right].\label{eq:app-whatwewanttocompute}
\end{align}
The fraction including the trigonometric function may seem awkward but stems from the fact that we redefined $\vb F\mapsto \vb g$ by scaling out only the magnitude $b=|\vb b|$. Had we instead used $\vb F=\vb b g$ with a scalar function $g$ (as in Ref.~\cite{Aurenche:2002wq, Moore:2021jwe}), the trigonometric function would not have appeared in \eqref{eq:app-whatwewanttocompute}. The definition used here is more convenient for obtaining the angular information discussed in the following section.

\subsection{Angular information}
For the angular information, the function $\vb g$ and the effective potential $D(\vb b)$ are decomposed into Fourier modes,
\begin{align}
    \vb g(b,\phi_b)&=\sum_n\vb g_n(b)e^{in\phi_b}, \\
    D(b,\phi_b)&=\sum_m D_m(b)e^{im\phi_b},
\end{align}
where $D_m(b)$ can be computed via
\begin{align}
    D_m(b)&=\frac{1}{2\pi}\int_0^{2\pi}\dd{\phi_b}e^{-im\phi_b}D(b,\phi_b).
\end{align}
In terms of these Fourier modes $\vb g_n$, the differential equation becomes
\begin{align}
    \left[\frac{A}{B}-\partial_b^2-\frac{3}{b}\partial_b+\frac{n^2-1}{b^2}\right]\vb g_n(b)=\sum_m\frac{D_m(b)}{B}\vb g_{n-m}(b).\label{eq:equation_gn}
\end{align}
Note that the boundary condition \eqref{eq:boundarycondition} only affects the modes $n=\pm1$, which 
can be seen by rewriting Eq.~\eqref{eq:boundarycondition} using exponential functions,
\begin{align}
	\lim_{b\to 0}\vb g(b,\phi_b)=\frac{1}{2B\pi}\frac{1}{b^2}\begin{pmatrix}
	    i(e^{i\phi_b}+e^{-i\phi_b})\\e^{i\phi_b}-e^{-i\phi_b}
	\end{pmatrix}.
    \label{eq:boundarycondition-exponentials}
\end{align}
This fixes the small-$b$ limit of $\vb g_{\pm 1}$.
In the isotropic limit, $D_m\sim \delta_{m0}$ and different Fourier modes decouple. Then,
only the modes for $m=\pm 1$ contribute to the rate. This isotropic case will be discussed in more detail in Section~\ref{sec:special-case-isotropic}.

Let us now consider the case of small $b$, where $D\sim b^2\log b$ can be neglected against $A$, i.e., the region where $|D_m(b)|\ll |A|$. In this region, Eq.~\eqref{eq:equation_gn} exhibits an analytic solution,
\begin{align}
	\vb g_n(b)=\frac{\vb c_1 I_{|n|}(b\sqrt{A/B}) + \vb c_2 K_{|n|}(b\sqrt{A/B})}{b},
\end{align}
where $I_n$ and $K_n$ are the modified Bessel functions of the first and second kind.
Thus, at small $b$, the general solution is given by the linear combination
\begin{align}
	\vb g(b,\phi_b)=\sum_n e^{in\phi_b} \frac{\vb c_I^n I_{|n|}(b\sqrt{A/B}) + \vb c_K^n K_{|n|}(b\sqrt{A/B})}{b}.\label{eq:general-solution}
\end{align}
In practice, the series will need to be truncated, and it will be enough to only consider modes with $-\nmax \leq  n \leq \nmax$, leading to $n_{\mathrm{fourier}}=2n_{\mathrm{max}}+1$ different Fourier modes for every component of $\vb g = (g_x, g_y)$. This results in $2(2n_{\mathrm{max}}+1)$ linearly independent solutions at small $b$. 
Eq.~\eqref{eq:equation_gn} is a coupled system of $2n_{\mathrm{max}}+1$ ordinary second-order differential equations, which requires $2(2n_{\mathrm{max}}+1)$ boundary conditions, which fix all the $c_I^n$ and $c_K^n$ uniquely. One natural boundary condition is to impose regularity at infinity \cite{Aurenche:2002wq},
\begin{align}
	\lim_{b\to\infty}\vb g_n(b)=0,\label{eq:boundary_cond_infinity}
\end{align} 
which yields $2n_{\mathrm{max}}+1$ conditions for every component of $\vb g$.
Another boundary condition is given by Eq.~\eqref{eq:boundarycondition-exponentials} at small $b$. To achieve that, it is useful to expand the Bessel functions for small $b$, 
\begin{align}
 I_n(b)/b &\sim b^{n-1}(1+\mathcal O(b)), \\ K_n(b)/b&\sim b^{n-1}(1+\mathcal O(b)) + \# b^{-n-1}(1+\mathcal O(b))\label{eq:Kn_expansion},
\end{align}
where the $\#$-symbol in \eqref{eq:Kn_expansion} denotes a possibly different proportionality constant than the first term. Which of the two terms in \eqref{eq:Kn_expansion} dominates depends on the value of $n$.

This fixes $c_K^{\pm 1}$ (2 additional conditions per component of $\vb g$). Furthermore, no function may diverge more quickly than $1/b^2$ at the origin, which fixes $\vb c_K^m=0$ for $m\geq 2$, resulting in $2(n_{\mathrm{max}}-1)$ additional conditions per component of $\vb g$.
This leaves still one missing (complex) boundary condition to determine the system completely.\footnote{One might wonder if excluding the $n=0$ modes would solve the problem: This would reduce the number of independent solutions by $2$, and the system would then be overdetermined.}
Since for the rate only the imaginary part of the constant to which $\vb g_n$ converges (see Eq.~\eqref{eq:app-whatwewanttocompute}) for $b\to 0$ contributes, and since both $K_0/b$ and $I_0/b$ diverge in that limit, one complex (or two real) boundary condition can be used to set $\IM \,\vb c_I^0=\IM\, \vb c_K^0=0$. To summarize, the boundary conditions are given by
\begin{subequations}\label{eq:app-boundaryconditions-combined}
\begin{align}
    \lim_{b\to\infty}\vb g_n(b)&=\vb 0,\label{eq:boundary_cond_infinity2}\\
	\vb c_K^1&=\frac{\sqrt{A/B}}{2B\pi}\begin{pmatrix}
	    i\\1
	\end{pmatrix},\\
    \vb c_K^{-1}&=\frac{\sqrt{A/B}}{2B\pi}\begin{pmatrix}
	    i\\-1
	\end{pmatrix},\\
    \IM\,\vb c_I^0&=\vb 0,\\
    \IM\,\vb c_K^0 &= \vb 0,
\end{align}
\end{subequations}
where the $\vb c_I$ and $\vb c_K$ coefficients are given by the small-$b$ approximation of the full solution.

The differential equation can be solved with these boundary conditions by using the fact that it is a linear homogeneous equation, and thus any linear combination of solutions also solves the equation. Hence, the $m$ linearly independent systems $\{\vb g_n^{(m)}\}$ are solved with $m$ independent initial conditions
\begin{align}
	\vb c_K^n{}^{(m)}&=0, & |n| \geq 2. \label{eq:initialcondition-enforcement}
\end{align}
Every system is initialized with exactly one nonzero coefficient.
This leads to $2(2n_{\mathrm{max}}+1) - 2(n_{\mathrm{max}}-2)=2\nmax + 6$ linearly independent sets of solutions $\{\vb g_n^{(m)}\}$. The full solution may be obtained by superimposing
\begin{align}
	\vb g(b,\phi_b)=\sum_{m}\vb a_m  g^{(m)}(b,\phi_b),
\end{align}
by choosing the coefficients $\vb a_m$ such that the boundary condition \eqref{eq:app-boundaryconditions-combined} is fulfilled.
It is convenient to put the information of the vector components of $\vb g$ into the coefficients $\vb a_m$. The advantage of that is that the $m$ systems need to be solved only once, and then two different sets of coefficients $\{a_m^x, a_m^y\}$ are obtained afterwards for the $x$ and $y$ solution.
In practice, this leads to
a linear system for the coefficients $\vb a_m$,
\begin{subequations}\label{eq:linearsystem-complete}
\begin{align}
    \sum_m \vb a_m g_n^{(m)}(\bmax)&=\vb {0}, \label{eq:boundary-condition-vanishing-infinity}\\
    \sum_m \vb a_m c_K^1{}^{(m)}&=\frac{\sqrt{A/B}}{2B\pi}\begin{pmatrix}
	    i\\1
	\end{pmatrix},\label{eq:actual_boundary-condition1}\\
    \sum_m\vb a_m  c_K^{-1}{}^{(m)}&=\frac{\sqrt{A/B}}{2B\pi}\begin{pmatrix}
	    i\\-1
	\end{pmatrix},\label{eq:actual_boundary-condition2}\\
    \sum_m  \IM\,\vb a_m c_I^0{}^{(m)}&=\vb 0,\\
    \sum_m \IM\,\vb a_m c_K^0{}^{(m)} &= \vb 0.
\end{align}
\end{subequations}
I have checked explicitly that the results in this paper are independent of the precise value chosen for $\bmax$.

Every system with index $(m)$ is initialized at a value $b=b_{\mathrm{min}}$, chosen such that $D(b_{\mathrm{min}},\phi)/A<0.00001$ with exactly one coefficient $c_{I,K}^n$ (taken to be a scalar, not a vector) in Eq.~\eqref{eq:general-solution} nonzero (except for those in Eq.~\eqref{eq:initialcondition-enforcement}). For every system $(m)$, the system \eqref{eq:equation_gn} is then integrated outwards using a fourth-fifth order Runge Kutta with adaptive time stepping until the absolute value of one of the solutions becomes larger than a threshold.
Of all the systems, the smallest of the maximum $b$ is taken, and the linear system \eqref{eq:linearsystem-complete} is solved.
In practice, for larger $\nmax$, the linear system becomes increasingly difficult to solve numerically,
but for the cases considered here, it was enough to take $\nmax=3$.
Finally, the solution for small $b$ can be written as
\begin{align}
\begin{split}
	\vb g(b,\phi_b)&=\sum_m \vb a_m\\
    &\times\sum_n e^{in\phi_b} \frac{c_I^n{}^{(m)} I_{|n|}(b\sqrt{A/B}) + c_K^n{}^{(m)} K_{|n|}(b\sqrt{A/B})}{b},\label{eq:general-solution_finally}
    \end{split}
\end{align}
where $c_{I,K}^n{}^{(m)}$ are the initial conditions.

Eventually, the prefactor of the $I_1$-solution is needed, because in the limit $b\to 0$, only the $I_1(b\sqrt{A/B})=\frac{\sqrt{A/B}}{2}+\mathcal O(b^2)$ contribute. Thus, at small $b$,
\begin{align}
    \vb g(0,\phi_b) = \sum_m\vb a_m \frac{\sqrt{A/B}}{2}\left(c_I^{1(m)}e^{i\phi_b}+c_I^{-1(m)}e^{-i\phi_b}\right).
\end{align}
Comparing with \eqref{eq:app-whatwewanttocompute}, this seemingly places additional requirements\footnote{These requirements amount to the full function $g(\phi_b)$ to be odd or even.} on the coefficients $(a_m)_x$ and $(a_m)_y$.
However, as discussed in Appendix \ref{sec:symmetry}, these additional conditions are actually a consequence of the system \eqref{eq:linearsystem-complete}.
This leads to
\begin{subequations}\label{eq:actually-getting-the-coefficients}
\begin{align}
    \IM \, g_x(0)/\cos\phi_b&=\sqrt{A/B}\,\IM\sum_m (a_m)_x c_I^{1(m)}\\
    &=\sqrt{A/B}\,\IM\,\sum_m(a_m)_xc_I^{-1(m)},\\
    \IM \, g_y(0)/\sin\phi_b&=\sqrt{A/B}\,\IM\left(i\sum_m (a_m)_x c_I^{1(m)}\right)\\
    &=-\sqrt{A/B}\,\IM \left(i\sum_m(a_m)_xc_I^{-1(m)}\right).
\end{align}
\end{subequations}

\subsection{The special case of an isotropic kernel\label{sec:special-case-isotropic}}
In an isotropic system, the potential $D_m$ has only an $m=0$ mode, leaving the different Fourier modes in Eq.~\eqref{eq:equation_gn} uncoupled. Then, only the $n=\pm 1$ modes have to be solved, leading to the equation
\begin{align}
        \left[\frac{A}{B}-\partial_b^2-\frac{3}{b}\partial_b\right]\vb g(b)=\frac{D(b)}{B}\vb g(b).\label{eq:equation_g_iso}
\end{align}
The boundary conditions are given by
\begin{align}
    \lim_{b\to\infty}\vb g(b)=0, && \lim_{b\to 0}\vb g(b)=\frac{i}{B\pi b^2}\begin{pmatrix}
        \cos\phi_b\\\sin\phi_b
    \end{pmatrix}.
\end{align}
In practice, these boundary conditions are enforced by solving the differential equation \eqref{eq:equation_g_iso} twice with the initial conditions
\begin{align}
    g^{(1)}(\bmin)=c_1\frac{I_1(b\sqrt{A/B})}{b}, && g^{(2)}(\bmin)=c_2\frac{K_1(b\sqrt{A/B})}{b},
\end{align}
with $c_i\in\mathbb C$. 
The solution of the homogeneous equation \eqref{eq:equation_g_iso} satisfying the boundary conditions is then obtained as the linear combination
\begin{align}
    g(b)=a_1g^{(1)}(b)+a_2g^{(2)}(b)
\end{align}
such that 
\begin{align}
    g(\bmax)=0,
\end{align}
leading to
\begin{align}
    \IM \frac{g_x(0)}{\cos\phi_b}=\IM\left(-i\frac{\sqrt{A/B}}{\pi B}\frac{g^{(2)}(\bmax)}{g^{(1)}(\bmax)}\right)\frac{\sqrt{A/B}}{2}.
\end{align}

\section{Numerical results}
\begin{figure}
    \centering
    \includegraphics[width=\linewidth]{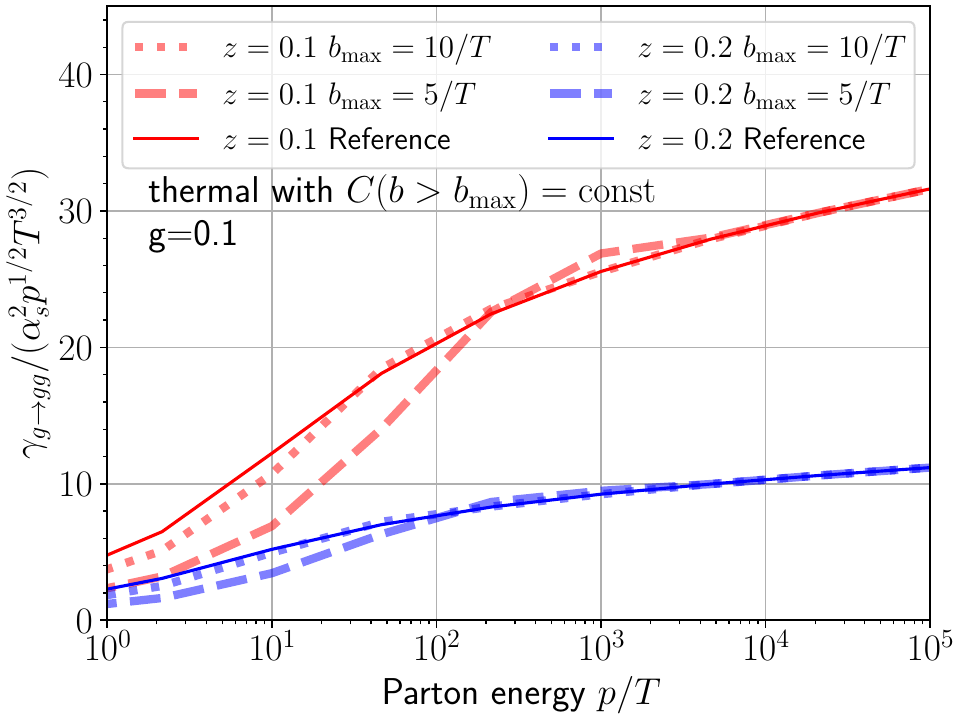}
    \includegraphics[width=\linewidth]{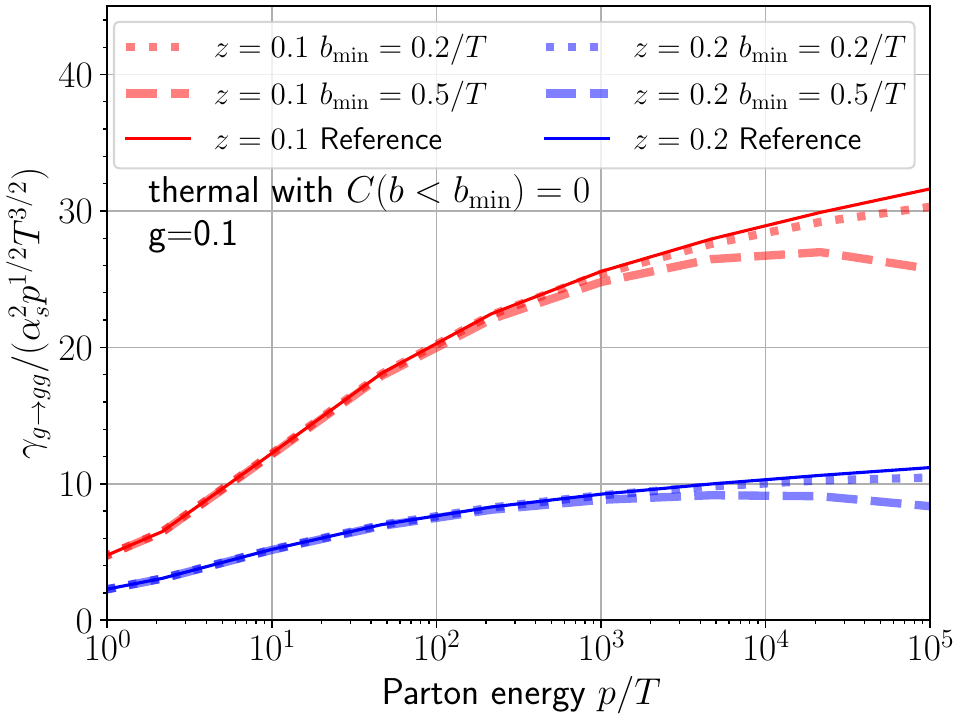}
    \caption{
    Gluon splitting rate $\gamma$ for an isotropic thermal collision kernel compared with modifications at large (top panel) and small $b$ (bottom panel).
    }
    \label{fig:rate_isotropic_crosscheck}
\end{figure}
We will now move on to discuss the numerical results for the splitting rates, first obtained for an isotropic collision kernel, and then using the method described in the previous section for the anisotropic model described in Section~\ref{sec:collkern-equ-anisotropic}.
For the numerical results, the numerical value $g=0.1$ is used for the coupling.

\subsection{Isotropic distributions}
As a numerical crosscheck, the rate is calculated for an isotropic dipole cross section \eqref{eq:Cb_eq_appr},
which is modified at large (Fig.~\ref{fig:rate_isotropic_crosscheck} top panel) or small distances (Fig.~\ref{fig:rate_isotropic_crosscheck} lower panel).
Both panels of \fig\ref{fig:rate_isotropic_crosscheck} depict the rate obtained from Eq.~\eqref{eq:Cb_eq_appr} as solid curve, with different colors denoting different splitting fractions $z$. In the top panel, the large-$b$ behavior is modified such that it remains constant for $b>\bmax$. Different line styles denote different values of $\bmax$, where for smaller values more modifications are visible. This demonstrates that the large-$b$ behavior of the dipole cross section is relevant for small parton energies. Different values of $z$ lead to qualitatively similar behavior.

Conversely, for the lower panel, the small-$b$ behavior is modified, such that $C(b<\bmin)=0$, with different line styles denoting different $\bmin$. Again, a larger value of $\bmin$ leads to larger effects on the rate, even changing the monotony of the curve at large parton energies.
This shows that for highly-energetic partons---which is relevant for jet quenching---the small-$b$ behavior of the dipole cross section is most relevant.

\subsection{Anisotropic systems}
\begin{figure*}
    \centering
    \centerline{
        \includegraphics[width=0.5\linewidth]{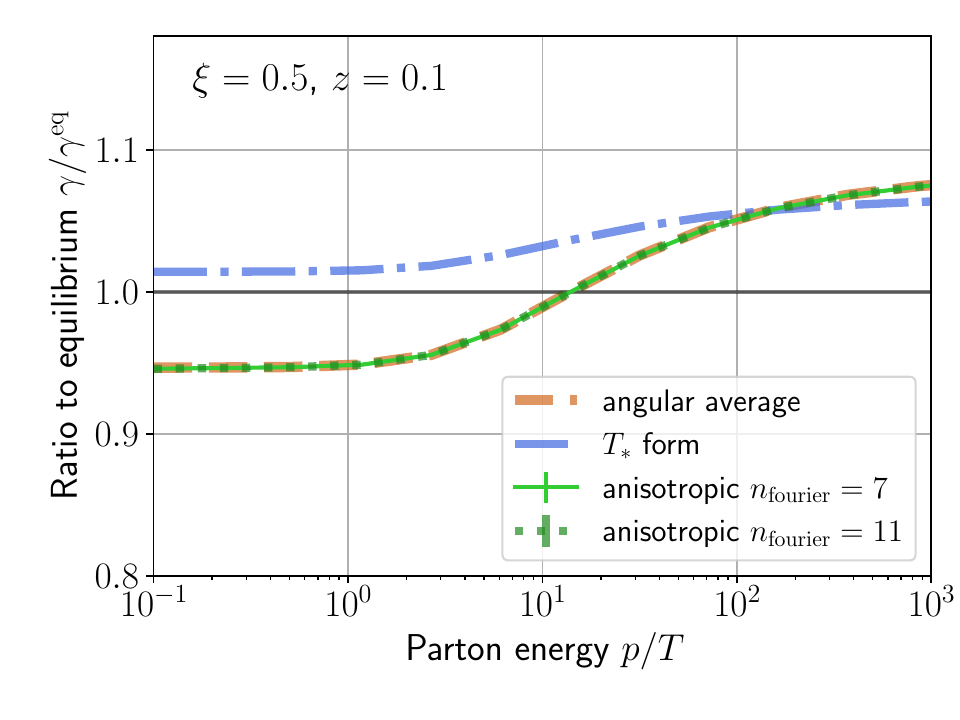}
        \includegraphics[width=0.5\linewidth]{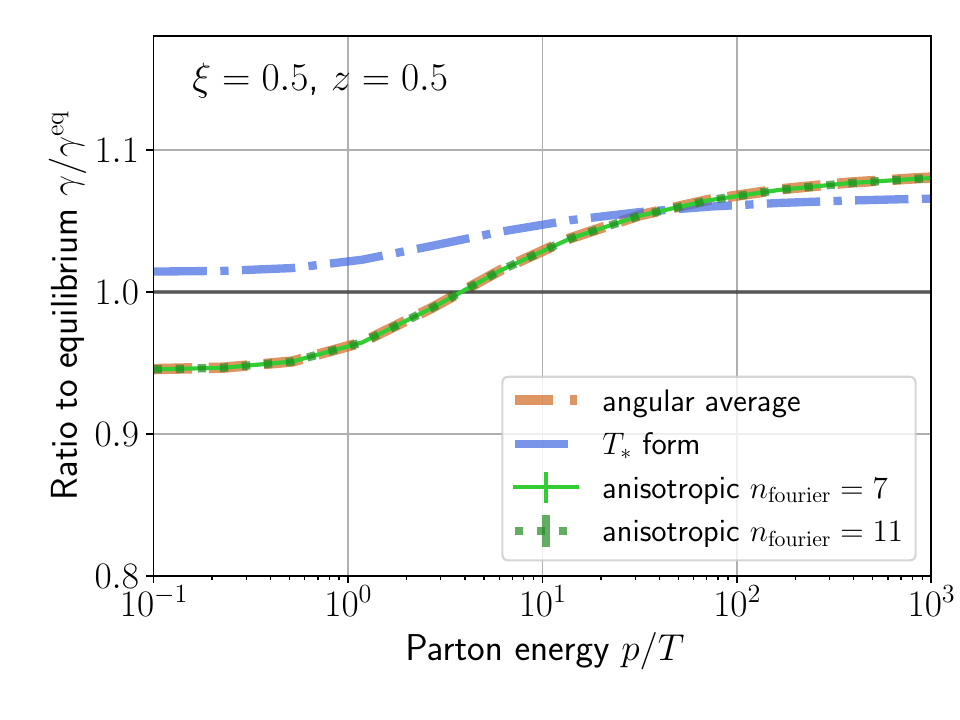}
    }
    \centerline{
        \includegraphics[width=0.5\linewidth]{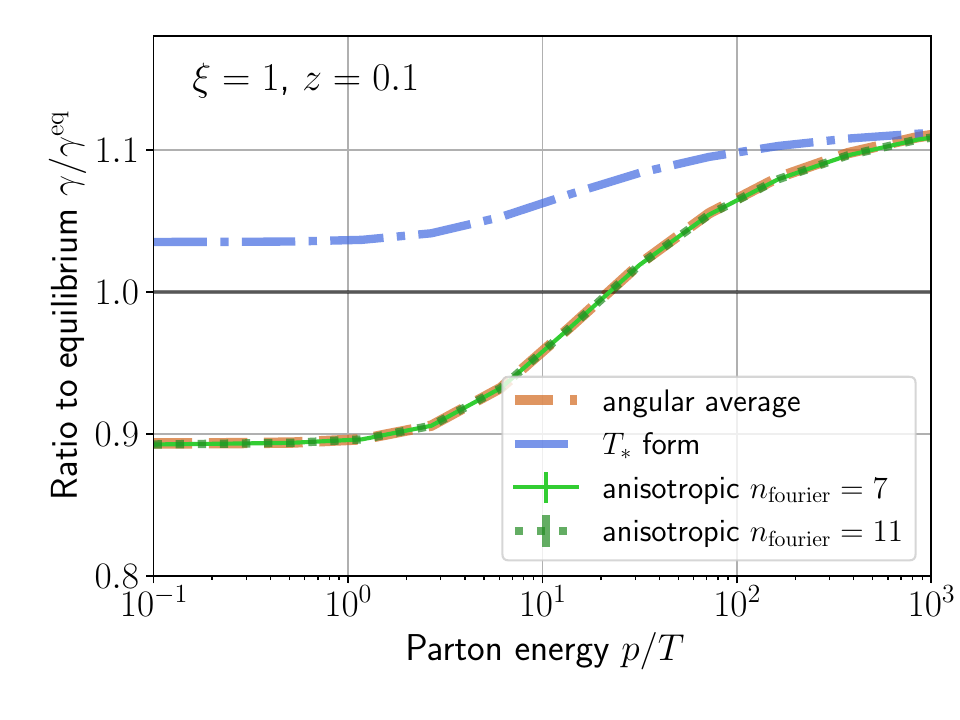}
        \includegraphics[width=0.5\linewidth]{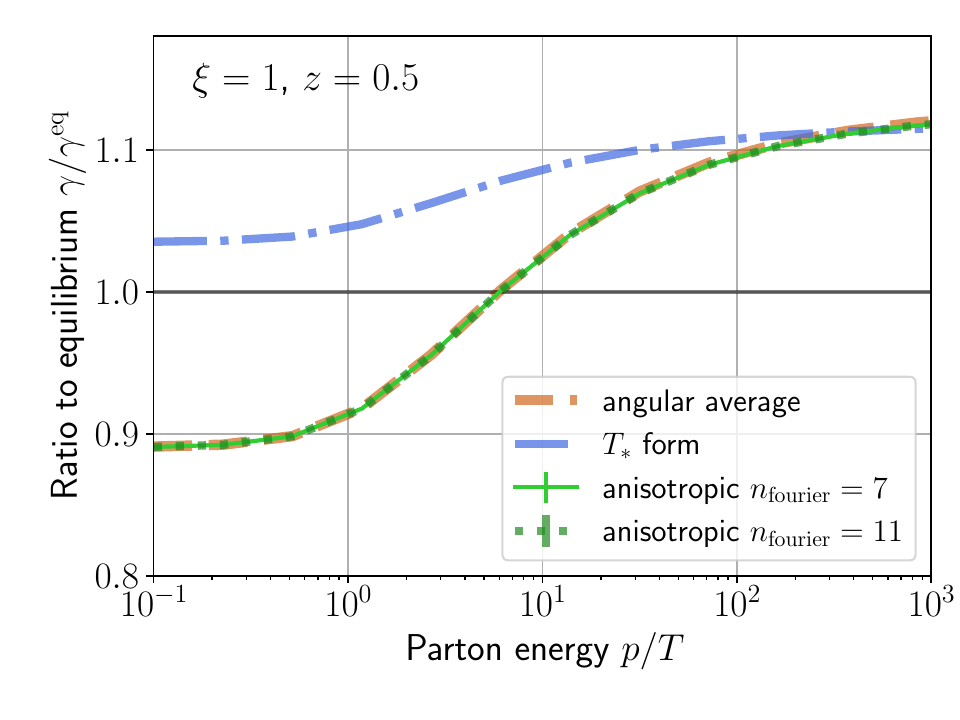}
    }
    \centerline{
        \includegraphics[width=0.5\linewidth]{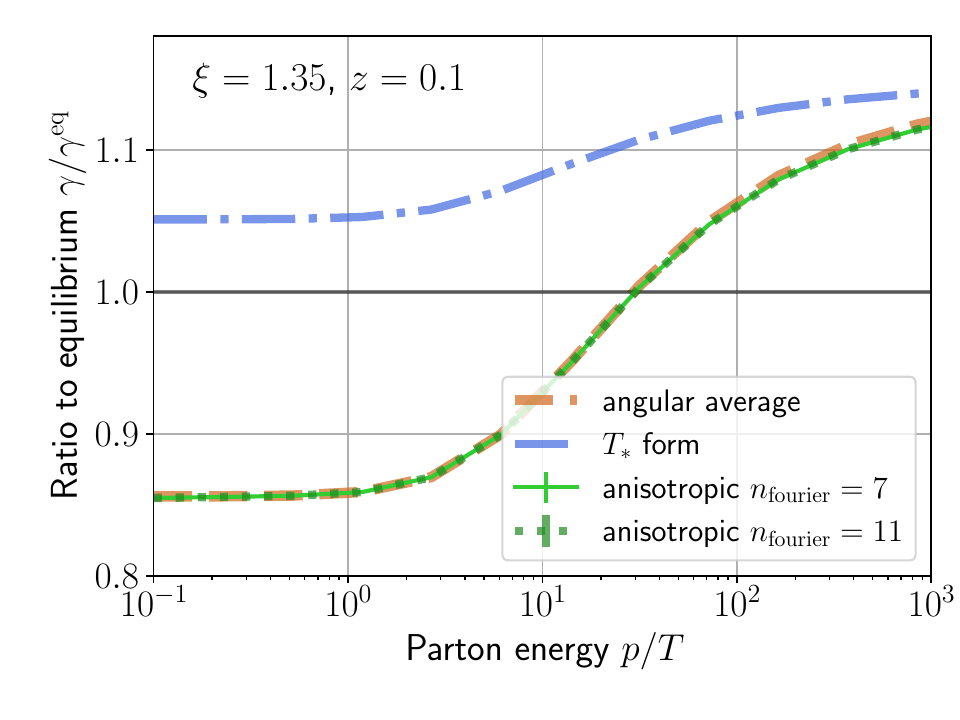}
        \includegraphics[width=0.5\linewidth]{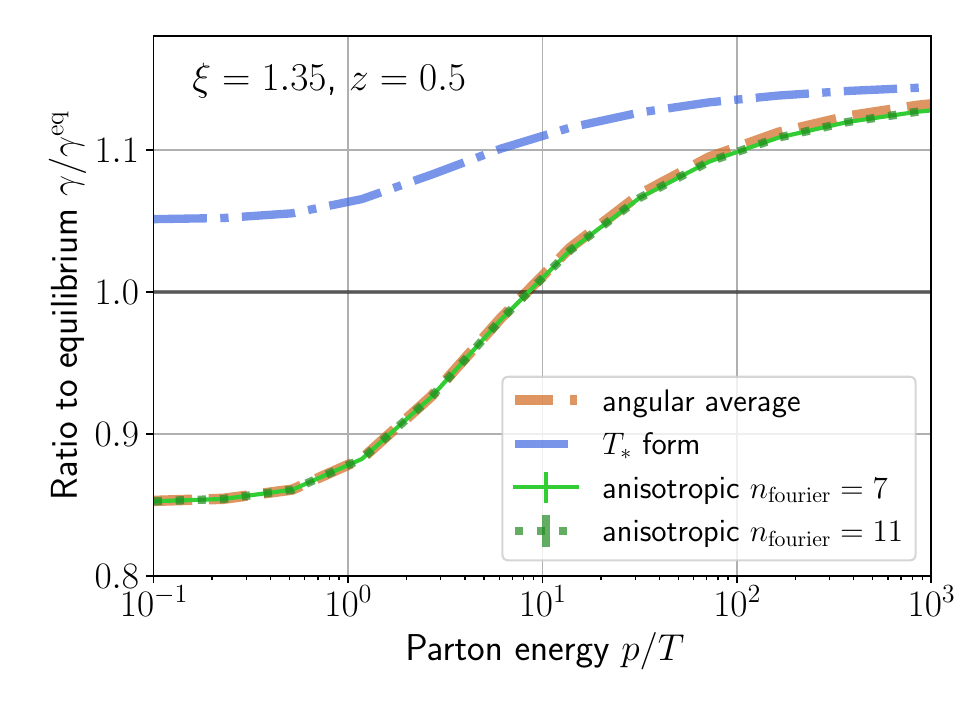}
    }
    \caption{Gluon splitting rate for an anisotropic dipole cross section (green curves) normalized to the thermal rate. The orange dashed line shows the rate obtained from the angular-averaged collision kernel. The blue dash-dotted line depicts the rate obtained from the isotropic approximated kernel \eqref{eq:Cb_iso_appr}.
    }
    \label{fig:rate_anisotropic_crosscheck}
\end{figure*}

\begin{figure*}
    \centering
    \centerline{
        \includegraphics[width=0.5\linewidth]{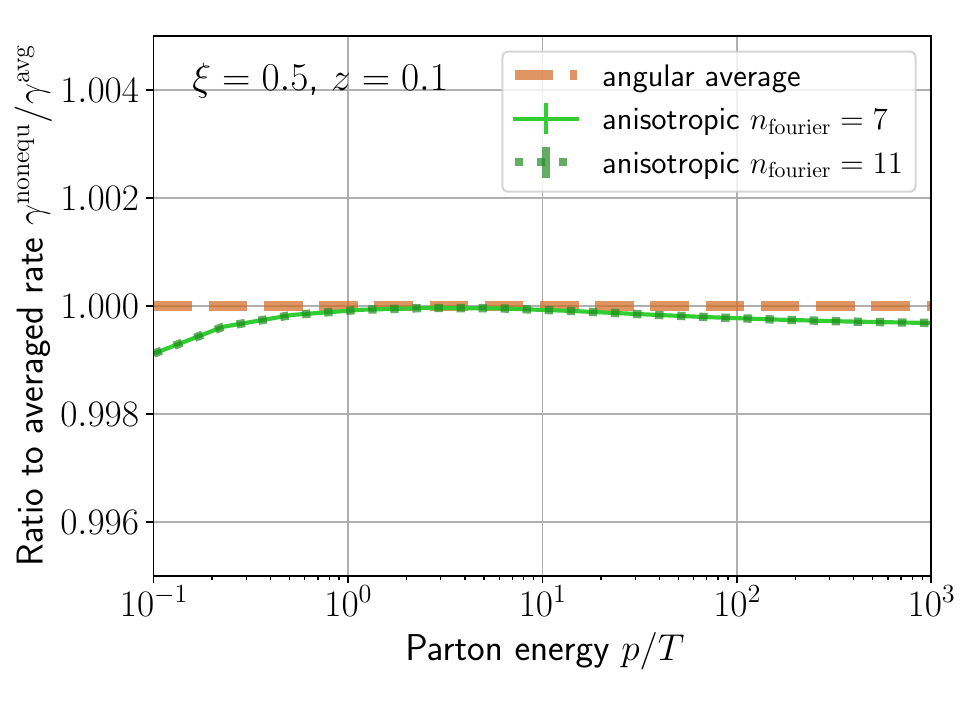}
        \includegraphics[width=0.5\linewidth]{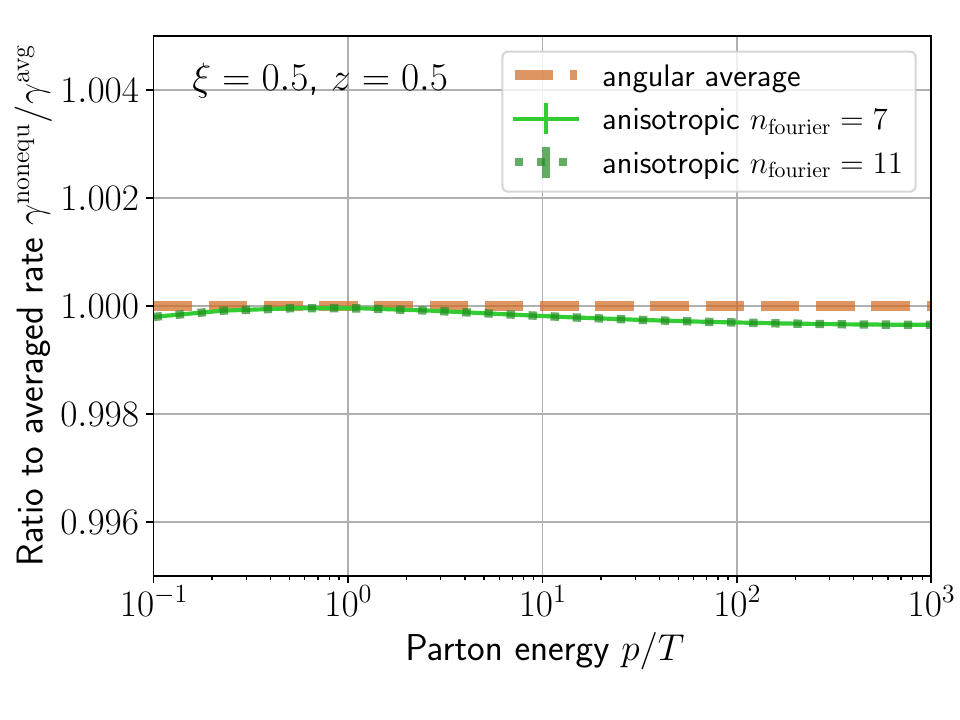}
    }
    \centerline{
        \includegraphics[width=0.5\linewidth]{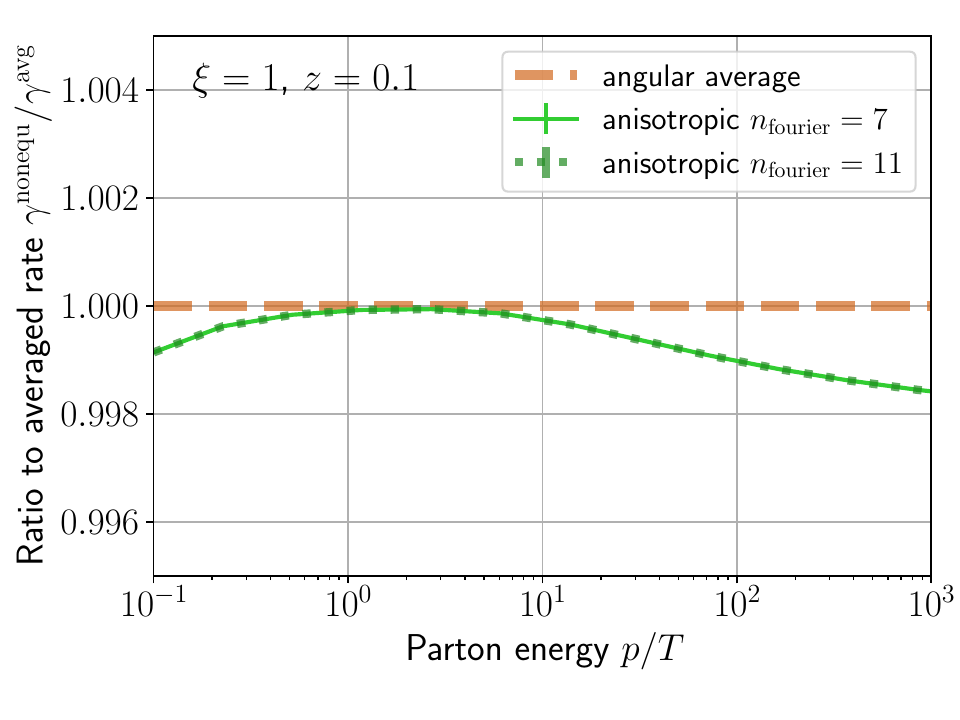}
        \includegraphics[width=0.5\linewidth]{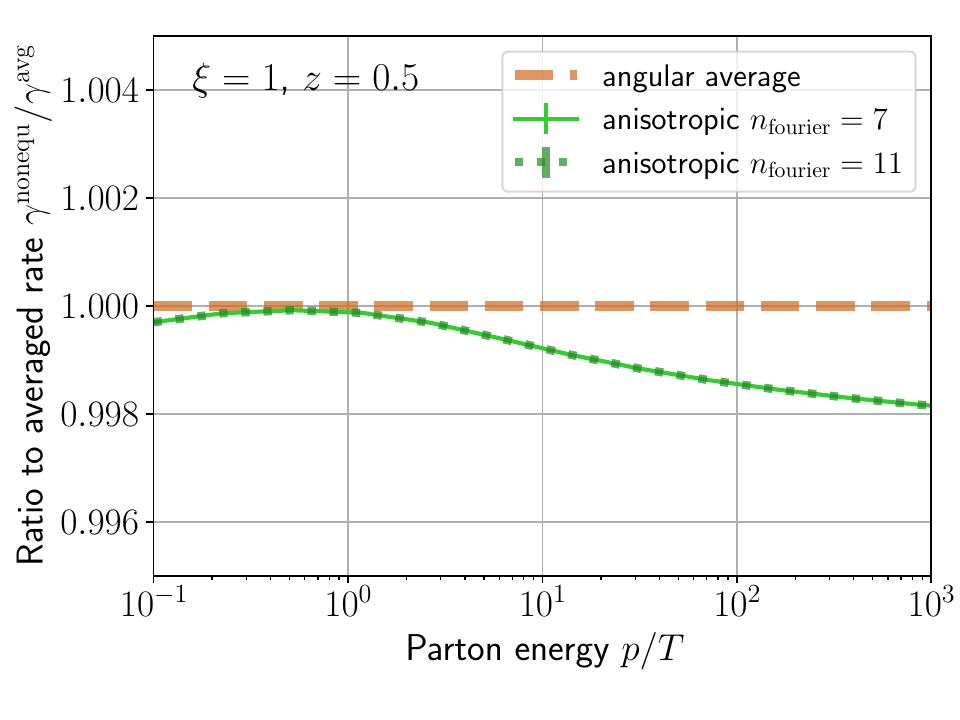}
    }
    \centerline{
        \includegraphics[width=0.5\linewidth]{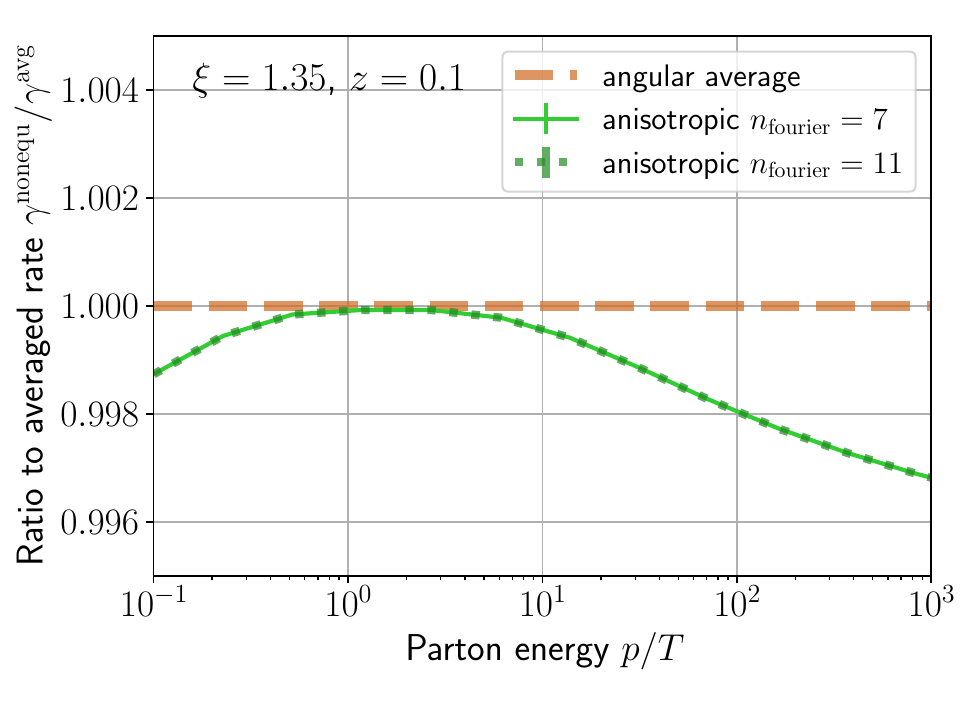}
        \includegraphics[width=0.5\linewidth]{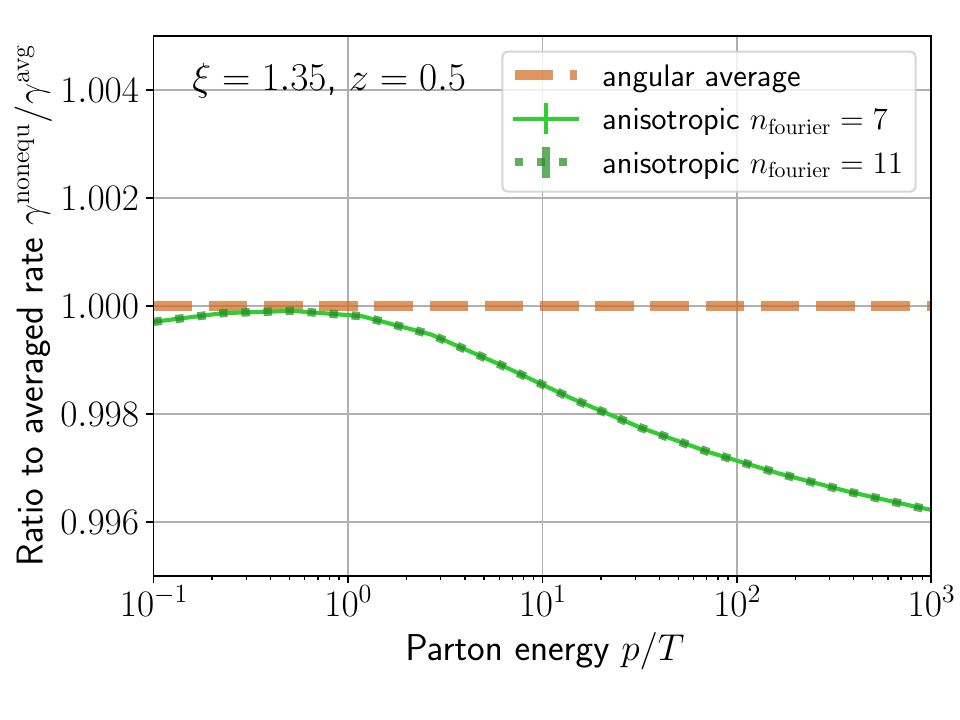}
    }
    \caption{
    Gluon splitting rate for an anisotropic dipole cross section (green curves) normalized to the rate obtained from the angular-averaged kernel.
    }
    \label{fig:rate_anisotropic_crosscheck2}
\end{figure*}
Let us now move on to discuss the results for the rates obtained from the anisotropic collision kernel introduced in Section \ref{sec:collkern-equ-anisotropic}, which are shown in Figs.~\ref{fig:rate_anisotropic_crosscheck} and \ref{fig:rate_anisotropic_crosscheck2}.
Both figures show different anisotropy parameters $\xi=0.5$ in the top, $\xi=1$ in the center, and $\xi=1.35$ in the lower panels. The left panels show the splitting ratio $z=0.1$, while the right panels show $z=0.5$.

Fig.~\ref{fig:rate_anisotropic_crosscheck} shows the rate as a function of the initial gluon energy, normalized to the equilibrium rate. The nonequilibrium rates obtained from the anisotropic kernel are shown as solid and dotted green lines, the rate obtained from the angular averaged kernel
\begin{align}
    \langle C(b)\rangle_{\phi_b}=\int_0^{2\pi}\frac{\dd{\phi_b}}{2\pi}C(b,\phi_b)
\end{align}
as dashed orange lines. Remarkably, the angular-averaged kernel provides a very good approximation of the anisotropic kernel, as can be seen from the good overlap of the orange and green lines. Their difference will be discussed below (shown in Fig.~\ref{fig:rate_anisotropic_crosscheck2}).

The isotropic approximated form using the effective infrared temperature $T_\ast$ from Eq.~\eqref{eq:tstar-calculated} and the nonequilibrium Debye mass $m_D$ from \eqref{eq:debyemass-calculated} differs from the nonequilibrium rate by more than $10\%$, 
with the largest deviations at smaller parton energies around the scale $T$.
However, this approximation is widely used in QCD kinetic theory simulations \cite{AbraaoYork:2014hbk, Kurkela:2014tea, Kurkela:2015qoa, Kurkela:2018oqw, Kurkela:2018xxd, Du:2020dvp, Du:2020zqg, kurkela_2023_10409474, Boguslavski:2024kbd, BarreraCabodevila:2025ogv} to obtain the collinear splitting rates, where the form \eqref{eq:Cb_iso_appr} is used to approximate the nonequilibrium kernel. In a companion paper \cite{Altenburger:2025iqa}, this is studied for more realistic kernels, showing even larger deviations.
This questions the accuracy of the current treatment of the collinear splitting rates in these simulations.

Finally, Fig.~\ref{fig:rate_anisotropic_crosscheck2} shows the ratio of the rate obtained from the anisotropic kernel and the angular averaged kernel. For all considered anisotropy parameters and splitting ratios, they differ by less than $0.5\%$. This demonstrates that the rate obtained from the angular averaged kernel provides a good approximation of the rate obtained from the anisotropic kernel. Varying the anisotropy parameter $\xi$ (different rows) reveals a continuous evolution, with larger deviations for larger anisotropies, and negligible deviations around the temperature scale $T$.

The two nonequilibrium green curves (solid and dotted) are obtained using a different number of Fourier modes. While the solid curve is obtained for $n_{\mathrm{fourier}}=7$, for the dotted curve $n_{\mathrm{fourier}}=11$ is used. They show excellent agreement, highlighting that $n_{\mathrm{fourier}}=7$, corresponding to $\nmax=3$, is sufficient for the parameter range considered in this work.

\section{Conclusion}
This work studies gluon splitting rates in an anisotropic gluon plasma using a novel numerical method.
The only medium input is the (anisotropic) dipole cross section $C(\vb b)$, related to the usual collision kernel $C(\vb q_\perp)$ via a Fourier transform (see Eq.~\eqref{eq:fouriertrafo}). The rates are obtained in the AMY formalism, which features several approximations: The collision kernel is assumed to remain constant during the splitting process, while the medium is assumed to have infinite extent. Despite these approximations, the rates obtained in this formalism are widely used in current QCD kinetic theory implementations.

The novel method presented here generalizes previous calculations by taking into account an anisotropic collision kernel $C(\vb q_\perp)$ using an expansion in Fourier modes. This leads to a drastically increased numerical complexity, as significantly larger systems of differential equations need to be solved numerically.
While in this paper, the method is applied to a simple anisotropic model for the collision kernel, it is also applicable to more general cases. In a companion paper \cite{Altenburger:2025iqa}, we study the rates obtained from a more realistic collision kernel.

The numerical results obtained in the present paper reveal that calculating the rate using an angular-averaged (isotropic) collision kernel $\langle C(b)\rangle_{\phi_b}$ provides a remarkably good approximation, with sub-percent differences to the result from the anisotropic kernel.
Nevertheless, the rate still substantially differs from its equilibrium form or from common approximations used in QCD kinetic theory implementations.
It will be interesting to study the impact of these results on QCD kinetic theory simulations.

Finally, it should be noted that the fact that the angular-averaged collision kernel provides a good approximation of the nonequilibrium kernel may be observable-dependent. In this paper, the splitting rate depends only on the energy of the emitted gluon, and its angular information and transverse momentum extent is integrated out. More differential observables, while being significantly more advanced to compute (e.g., differential in the transverse momentum \cite{Apolinario:2014csa}), may be more sensitive to the anisotropies present in the plasma \cite{Hauksson:2023tze, Barata:2024bqp}.
Nevertheless, the numerical method introduced in this work may provide a starting point for more advanced, differential studies, possibly enabling the identification of
experimental observables that probe the initial stages of heavy-ion collisions, where the QCD plasma is still far from equilibrium.

\begin{acknowledgements}
    I would like to thank Kirill Boguslavski for insightful discussions about the numerical method introduced here, for convincing me to write it up in a paper, and for useful comments regarding the manuscript.

    FL is a recipient of a DOC Fellowship of the Austrian Academy of Sciences at TU Wien (project 27203). This work is funded in part by the Austrian Science Fund (FWF) under Grant DOI 10.55776/P34455, and Grant DOI 10.55776/J4902.
    For the purpose of open access, the author has applied a CC BY public copyright license to any Author Accepted Manuscript (AAM) version arising from this submission.

\end{acknowledgements}

\appendix
\section{Delta function imposes boundary condition\label{app:delta}}
This appendix demonstrates that Eq.~\eqref{eq:boundarycondition} is the solution to the differential equation Eq.~\eqref{eq:differentialequation-delta}, which reads
\begin{align}
	-2i\nabla\delta(\vb b)=-B\nabla^2 F(\vb b).\label{eq:differentialequation-delta-app}
\end{align}
It can be solved by going to Fourier space, where we may use the integral representation of the delta function,
\begin{align}
	\delta (\vb b)=\int\frac{\dd[2]{\vb q_\perp}}{(2\pi)^2}e^{i\vb q_\perp\cdot \vb b},
\end{align}
such that $\vb F(q_\perp) = 2\frac{\vb q_\perp}{B q_\perp^2}$. The backward Fourier transform can be done using the parameterization $\vb b = b(\cos\phi_b,\sin\phi_b)$, $\vb q_\perp=q_\perp(\cos\phi_q,\sin\phi_q)$,
\begin{subequations}
\begin{align*}
	\int\dd[2]{\vb q_\perp}&\frac{\vb q_\perp}{q_\perp^2}e^{i\vb b\cdot\vb q_\perp}\\
	&=b \int_0^{2\pi}\!\!\!\dd\phi_q\int_0^\infty\!\!\!\dd q_\perp \begin{pmatrix}
		\cos\phi_q\\\sin\phi_q
	\end{pmatrix} e^{ib q_\perp \cos(\phi_q-\phi_b)}\\
	&=b\int_0^{2\pi}\!\!\!\dd{\tilde\phi_q}\int_0^\infty\!\!\!\dd{q_\perp}\begin{pmatrix}
		\cos(\tilde \phi_q+\phi_b)\\ \sin(\tilde\phi_q + \phi_b)
	\end{pmatrix} e^{ib q_\perp \cos\tilde\phi_q}
	\\&=b\int_0^{2\pi}\!\!\!\dd{\tilde\phi_q}\int_0^\infty\!\!\!\dd{q_\perp}\\
	&\quad\times \begin{pmatrix}
		\cos\phi_b\cos\tilde\phi_q-\sin\phi_b\sin\tilde\phi_q\\ \sin\phi_b\cos\tilde\phi_q+\cos\phi_b\sin\tilde\phi_q
	\end{pmatrix} e^{ib q_\perp \cos\tilde\phi_q},\\
	&=2\pi i \frac{\vb b}{b^2}
\end{align*}
\end{subequations}
where we used
\begin{align*}
	\int_0^{2\pi}\dd\phi \cos\phi\,\, e^{i b q_\perp \cos\phi}&=2\pi i J_1(q_\perp b), \\ \int_0^{2\pi}\dd\phi \sin\phi\,\, e^{i b q_\perp \cos\phi}&=0.
\end{align*}

Eq.~\eqref{eq:differentialequation-delta} (and the full differential equation \eqref{eq:app-impactparameterspaceequation2}) is thus solved by
\begin{align}
	\lim_{b\to 0}\vb F(\vb b)=\frac{i}{B\pi}\frac{\vb b}{ b^2}.
\end{align}
This can be taken as a boundary condition, and Eq.~\eqref{eq:app-impactparameterspaceequation2} is solved for $b>0$.

\section{Symmetry\label{sec:symmetry}}
This appendix shows that the seemingly additional requirements \eqref{eq:actually-getting-the-coefficients} on the coefficients
follow from the system \eqref{eq:linearsystem-complete}. To do that, we first consider as a toy model the simplified case of having only four sets (labeled by the index $m$) of two coupled equations, and only consider the modes $n=\pm1$. With that assumption, the function $g^{(m)}$ for one set of initial conditions $c_i^{(m)}$ can be written as
\begin{align}
\begin{split}
    g^{(m)}&=e^{i\phi}(c_1^{(m)}f_1(b)+c_2^{(m)}f_2(b))\\
    &+e^{-i\phi}(c_3^{(m)}f_1(b)+c_4^{(m)}f_2(b)).
    \end{split}
\end{align}
As initial conditions, we assume that we use $c_i^{(m)}=\delta_i^m$.
In that simplified case, the requirement \eqref{eq:boundary-condition-vanishing-infinity}, i.e., vanishing at infinity, leads to
\begin{align}
    \begin{split}\label{eq:linearsystem-sample}
    c_1 a_1+c_2 a_2+c_3 a_3+c_4 a_4&=0\\
    c_3 a_1 + c_4 a_2 + c_1 a_3 + c_2 a_4&=0, 
    \end{split}
\end{align}
where we have used that $g^{(1)}$ and $g^{(3)}$ are similar because they are both initialized with the $f_1$ function. As an additional boundary condition, we need to have $a_1=\pm a_3$, enforcing symmetric or antisymmetric boundary conditions (corresponding to the cosine or sine in \eqref{eq:boundarycondition-exponentials}). Additionally, we fix the value of one coefficient, e.g., $a_1=1$, coming from \eqref{eq:actual_boundary-condition1}. Thus, we have the additional conditions (mimicking \eqref{eq:actual_boundary-condition1} and \eqref{eq:actual_boundary-condition2})
\begin{align}
    a_3=\pm a_1, && a_1 = 1. \label{eq:toymodel-symmetricondition}
\end{align}
Inserting this into \eqref{eq:linearsystem-sample}, we can subtract (or add) those two equations to obtain
\begin{align}
    a_2(c_2 \mp c_4)+a_4(c_4\mp c_2)=0,
\end{align}
which leads to $a_2=\pm a_4$, i.e., the symmetry condition for \eqref{eq:toymodel-symmetricondition} enforces the same symmetry on the other coefficients $a_2$ and $a_4$.

While this holds when only considering two modes, $n=\pm 1$, this is also true when adding higher modes. For instance, let us now consider adding also the $n=\pm 2$ Fourier modes, such that the system~\eqref{eq:linearsystem-sample} then reads
\begin{subequations}
\begin{align}
    c_1 a_1 + c_2 a_2 + c_3 a_3+c_4 a_4 + c_5 a_5 + c_6 a_6 &= 0\\
    c_3 a_1 + c_4 a_2 + c_1 a_3 + c_2 a_4 + c_6 a_5 + c_5 a_6 &=0\\
    c_ 7 a_1 + c_8 a_2 + c_9 a_3+c_{10}a_4+c_{11}a_5+c_{12}a_6&=0\\
    c_9 a_1 + c_{10}a_2 + c_7 a_3 + c_8 a_4+c_{12} a_5 + c_{11} a_6&=0
\end{align}
\end{subequations}
which, with the condition \eqref{eq:toymodel-symmetricondition}, leads to
\begin{subequations} \label{eq:system2}
\begin{align}
    0&=a_2(c_2\mp c_4)+a_4(c_4\mp c_2)+a_5(c_5\mp c_6)+a_6(c_6\mp c_5)\\
    0&=a_2(c_8\mp c_{10})+a_4(c_{10}\mp c_8)+a_5(c_{11}\mp c_{12})+a_6 (c_{12}\mp c_ {11}).
\end{align}
\end{subequations}
Similar as before, upon eliminating $a_5$ and $a_6$, this results in
\begin{align}
    0&=a_2\left(\frac{c_2\mp c_4}{c_5\mp c_6}-\frac{c_8\mp c_{10}}{c_{11}\mp c_{12}}\right)+a_4\left(\frac{c_4\mp c_2}{c_5\mp c_6}-\frac{c_{10}\mp c_8}{c_{11}\mp c_{12}}\right),
\end{align}
again implying $a_2=\pm a_4$, and thus the same symmetry condition as for the other coefficients $a_1$ and $a_2$.

More compactly, the system \eqref{eq:system2} can be written as 
\begin{subequations}
\begin{align}
    0&=\tilde c_1(a_2\pm a_4)+\tilde c_2(a_5\pm a_6),\\
    0&=\tilde c_3(a_2\pm a_4)+\tilde c_4(a_5\pm a_6),
\end{align}
\end{subequations}
with $\tilde c_1=c_2\mp c_4$, $\tilde c_2=c_5\mp c_6$ and similarly for $\tilde c_3$ and $\tilde c_4$. Redefining now $\tilde a_1 = a_2\pm a_4$ and $\tilde a_2=a_5\pm a_6$, we obtain the system
\begin{align}
    0=\tilde c_1\tilde a_1+\tilde c_2\tilde a_1, && 0&=\tilde c_3\tilde a_1+\tilde c_4\tilde a_2,
\end{align}
which, if regular, has the solutions $\tilde a_1=\tilde a_2=0$, implying $a_2=\pm a_4$. This argument generalizes easily to higher Fourier modes, and thus shows that $a_2=\pm a_4$ is indeed a consequence of \eqref{eq:linearsystem-complete}.

Thus, the symmetry between $c_I^{1(m)}$ and $c_I^{-1(m)}$ in Eq.~\eqref{eq:actually-getting-the-coefficients} is not an additional input but a consequence of the linear system \eqref{eq:linearsystem-complete}.

\bibliography{bib}
\end{document}